\documentclass[journal=jctcce,manuscript=article]{achemso}

\usepackage{achemso}

\usepackage{hyperref}
\usepackage{amsmath}
\usepackage{bbold}
\usepackage{xcolor}


\author{Jia Chen}
\affiliation[UFL_Physics]{Department of Physics, University of Florida, Gainesville, FL 32611, USA}
\alsoaffiliation[QTP]{Quantum Theory Project,  University of Florida, Gainesville, FL 32611, USA}
\email{jiachen@ufl.edu}

\author{Hai-Ping Cheng}
\affiliation[UFL_Physics]{Department of Physics, University of Florida, Gainesville, FL 32611, USA}
\alsoaffiliation[QTP]{Quantum Theory Project,  University of Florida, Gainesville, FL 32611, USA}
\email{hping@ufl.edu}
\author{J. K.  Freericks}
\affiliation[Georgetown_Physics]{Department of Physics, Georgetown University, 37th St. and O St., NW, Washington, DC 20057, USA}
\email{James.Freericks@georgetown.edu}

\title{Low-Depth Unitary Coupled Cluster Theory for Quantum Computation}

\begin{document}


\begin{abstract}
The unitary coupled cluster (UCC) approximation is one of the more promising wave-function ans\"atze for electronic structure calculations on quantum computers via the variational quantum eigensolver algorithm. However, for large systems with many orbitals, the required number of UCC factors still leads to very deep quantum circuits, which can be challenging to implement. Based on the observation that most UCC amplitudes are  small for weakly correlated molecules, we devise an algorithm that employs a Taylor expansion in the small amplitudes, trading off circuit depth for extra measurements. Strong correlations can be taken into account by performing the expansion about a small set of UCC factors, which are treated exactly. Near equilibrium, the Taylor series expansion often works well without the need to include any exact factors; as the molecule is stretched and correlations increase, we find only a small number of factors need to be treated exactly.
\end{abstract}

\section{Introduction}
Quantum computing provides a new paradigm for manipulating information according to the laws of quantum mechanics; it is expected to provide an advantage over classical computation for some scientific problems.\cite{Mike_Ike} As envisioned by Richard Feynman, one of those problems is simulating quantum mechanical systems.\cite{Feynman1982} Focusing on molecular quantum systems, quantum chemistry is poised to be among the fields of study that could benefit from the developments of quantum computation. An example for how this can be achieved is given by the phase estimation algorithm  \cite{Daniel1997, Daniel1999} for computing energy eigenvalues and preparing energy eigenstates. This algorithm has been simulated and shown to work\cite{Aspuru-Guzik2005}; it will provide opportunities for solving problems that cannot be solved on classical computers. Unfortunately, phase estimation is quite challenging to implement, because it requires controlled time evolution of the system. Hence, it has extremely deep circuits if the time evolution of the molecule is treated exactly in a conventional basis for the Hamiltonian.

Current quantum hardware is limited due to noise and decoherence. This near-term hardware is called noisy intermediate-scale quantum (NISQ)\cite{Preskill2018}. They can only work with low-depth circuits on a modest number of qubits. This precludes direct application of many quantum algorithms, such as the phase-estimation algorithm. To be specific, we have to work with two constraints in the near term: (i) the number of qubits will remain relatively small  and (ii) the allowed circuit depth will remain low due to  gate fidelity and decoherence errors. To take advantage of existing and near-term quantum computers, the variational quantum eigensolver (VQE) has been proposed as a low-depth alternative to quantum phase estimation. \cite{Peruzzo2014} It is a hybrid quantum-classical method, and it shows great promise.\cite{Cao2019} VQE needs to be carried out both on quantum and classical computers: on the quantum computer, quantum states depending on a set of variational parameters are prepared, and the expectation value of the Hamiltonian is then measured. Next, that set of parameters is optimized on classical computers and the loop is repeated until converged. But, this approach suffers from the need for higher-depth circuits as the ansatz wavefunction becomes more complex and from the appearance of  barren plateaus in the optimization space (which is exacerbated by the noise and decoherence of NISQ machines). Our approach attempts to resolve both of these issues by using a quadratic expansion of the energy in terms of the variational amplitudes, which allows for significantly lower depth in the required circuits and an optimization that is performed entirely on the classical computer. But, it does so at the expense of requiring significantly more measurements.

Originally proposed as a wave-function ansatz for quantum chemistry about four decades ago,\cite{Kutzelnigg1977, Koch1981, Kutzelnigg1983} unitary coupled cluster theory (UCC) has gained renewed attention because, in its factorized form, it can serve as an efficient state preparation method for the VQE. Usually, the UCC ansatz starts from the Hartree-Fock (HF) reference state in the occupation number representation; then the different UCC factors are applied to the HF states in sequence; this excites the HF state by including terms from unoccupied orbitals, but it also includes de-excitation terms when the prepared state includes terms that can be de-excited by the next UCC factor. General speaking, the circuit depth for UCC state preparation is proportional to the number of excitations applied, especially if they are the same rank excitation. For example, a molecule with $N$ orbitals, has $N^4$ doubles excitations, which are usually the most important excitations to include in the ansatz. As a result, including all (or just a fraction of) all possible singles and doubles excitations already requires a circuit depth that is prohibitively high for large molecules, especially so on NISQ hardware.  

In this contribution, we show how to improve this situation, by requiring only a small subset of UCC factors to be applied to the reference state on the quantum computer. The insight behind this comes from the fact that most amplitudes $\theta$ have small absolute values. This suggests that they can be expanded in a Taylor series about $\theta=0$, truncated at quadratic order and then optimized. Doing this from the reference HF state is only viable when the molecule is weakly correlated, which holds predominantly near the equilibrium configuration bond distances and angles. But, how many UCC factors need to be treated exactly when carrying out this approach as the correlations are increased (due to stretching)? This is the question we address in this work. We find the number remains relatively small, implying that such an approach can enable more complex molecules to be treated on currently available NISQ machines. Similar ideas have been used in other contexts as well. A quadratic expansion about a density-matrix-renormalization group calculation was performed successfully for a carbon dimer\cite{carbon} and the approach was also investigated for a quantum-inspired algorithm using an ansatz that can be constructed solely from Clifford circuits, which can be easily simulated on a classical computer~\cite{clifford}. Our approach is similar to both of these earlier works, but has a number of differences as well.

\section{Theory and Method}

\subsection{Unitary Coupled Cluster Theory (UCC) in Factorized Form and Operator Identity for UCC Factors}

In unitary coupled cluster (UCC) theory, the trial wave-function is expressed in an exponential form, given by
\begin{align}
    |\Psi_{\textrm{UCC}}\rangle = \exp (\hat{\sigma}) |\Psi_0\rangle , 
\end{align}
where $|\Psi_0\rangle $ is the reference state and the operator $\hat{\sigma}$ is an anti-Hermitian combination of particle-hole excitation and de-excitation:
\begin{align}
    \hat{\sigma} &= \hat{T} - \hat{T}^{\dagger} ; \\
    \hat{T} &= \sum_i^{occ}\sum_a^{vir} \theta_i^a \hat{a}_a^{\dagger} \hat{a}_i +  \sum_{ij}^{occ}\sum_{ab}^{vir} \theta_{ij}^{ab} \hat{a}_{a}^{\dagger}\hat{a}_{b}^{\dagger} \hat{a}_{j}\hat{a}_{i} + \cdots ~.
\end{align}
Here, the rotation angles $\theta$ are the variational parameters. We use letters from the start of the alphabet $a$, $b$, $c,\ldots$ to denote the virtual orbitals, with respect to the reference state,  and letters from the middle of the alphabet $i$, $j$, $k,\ldots$ to denote the occupied orbitals in the reference state. To simplify the notation, we express a general $n$-fold excitation operator as $\hat{a}^{ab\dots}_{ij\dots} = \hat{a}^{\dagger}_{a}\hat{a}^{\dagger}_{b} \dots \hat{a}_j \hat{a}_i$ (with the corresponding de-excitation operator being its Hermitian conjugate). We work with a factorized form for the  UCC, which is given by
\begin{align}
    |\Psi_{\textrm{UCC}}\rangle = U_{\textrm{UCC}} |\Psi_0\rangle = \prod_k e^{\theta_k\sigma_k}
    =\prod_{ij\cdots}^{occ}\prod_{ab\cdots}^{vir} \exp [\theta_{ij\cdots}^{ab\cdots} (\hat{a}_{ij\cdots}^{ab\cdots}-\hat{a}_{ab\cdots}^{ij\cdots})] |\Psi_0\rangle . 
    \label{eq:ucc-fac}
\end{align}
The factorized form is generally different from an ansatz that puts all operators in one exponential. But, because we are doing a variational calculation and there is flexibility given by the needed chemical accuracy, this factorized form is usually sufficient to achieve chemical accuracy, if enough factors are included. Indeed, if factors are repeated, it can be used to approximate the original UCC ansatz via the Trotter product formula.

For a general UCC factor, we derived a general operator identity based on a hidden SU(2) algebra.\cite{evangelista,luogen,jia} It is
\begin{align}\label{operator_identity}
\begin{split}
& U^{a_1\dots a_n}_{i_1\dots i_n} = \exp[\theta(\hat{a}^{a_1\dots a_n}_{i_1\dots i_n}-\hat{a}^{i_1\dots i_n}_{a_1\dots a_n})] = 1 + \sin\theta (\hat{a}^{a_1\dots a_n}_{i_1\dots i_n}-\hat{a}^{i_1\dots i_n}_{a_1\dots a_n}) \\
&+(\cos\theta-1)[\hat{n}_{a_1}\dots\hat{n}_{a_n}(1-\hat{n}_{i_1})\dots(1-\hat{n}_{i_n}) +(1-\hat{n}_{a_1})\dots(1-\hat{n}_{a_n})\hat{n}_{i_1}\dots\hat{n}_{i_n}].
\end{split}
\end{align}
Here the number operators are of the form $\hat{n}=\hat{a}^\dagger\hat{a}$.
If rotation angle $\theta = 0$, the UCC factor becomes the identity: $U(0)=\mathbb{I}$.

We can take the derivative of Eq.~(\ref{operator_identity}):
\begin{align}\label{operator_identity_2}
\begin{split}
&\frac{d \hat{U}^{a_1\dots a_n}_{i_1\dots i_n} (\theta)}{d \theta} =  \cos\theta (\hat{a}^{a_1\dots a_n}_{i_1\dots i_n}-\hat{a}^{i_1\dots i_n}_{a_1\dots a_n}) \\
&-\sin\theta [\hat{n}_{a_1}\dots\hat{n}_{a_n}(1-\hat{n}_{i_1})\dots(1-\hat{n}_{i_n}) +(1-\hat{n}_{a_1})\dots(1-\hat{n}_{a_n})\hat{n}_{i_1}\dots\hat{n}_{i_n}].
\end{split}
\end{align}
When we evaluate the derivative at $\theta =0$, it becomes the operator term in the exponent, namely
\begin{align}\label{Ud0}
\frac{d \hat{U}^{a_1\dots a_n}_{i_1\dots i_n} (\theta)}{d \theta} |_{\theta=0} =   \hat{a}^{a_1\dots a_n}_{i_1\dots i_n}-\hat{a}^{i_1\dots i_n}_{a_1\dots a_n} = \hat{\sigma}^{a_1\ldots a_n}_{i_1\ldots i_n}.
\end{align}
The derivative of the wavefunction with respect to one rotation angle immediately follows as
\begin{align}\label{ucc_first_derivative}
   \frac{ d U_{\textrm{UCC}}}{d \theta_k} = \hat{U}_n\cdots\hat{U}_{k+1}\frac{d \hat{U}_k}{d \theta_k}\hat{U}_{k-1}\cdots\hat{U}_2\hat{U}_1 .
\end{align}

Evaluated at $\theta_k=0$, a simple result is obtained:
\begin{align}
   \frac{ d U_{\textrm{UCC}}}{d \theta_k} \Big{|}_{\vec{\theta} =0}= \frac{d \hat{U}_k}{d \theta_k} \Big{|}_{\theta_k=0} = \hat{\sigma}_k.
\end{align}

\subsection{Quadratic Angle Expansion}
A Taylor expansion of the expectation value of the energy around the point where all angles vanish  ($\vec{\theta}=0$) becomes
\begin{align}\label{tot_energy}
\langle \hat{H} (\vec{\theta})\rangle = \langle \hat{H} (0)\rangle + \sum_{k} b_k \theta_k + \frac{1}{2}\sum_{k,m} A_{km}\theta_k\theta_m + \mathcal{O}(\theta^3)
\end{align}
where,
\begin{align}\label{b_vec}
b_k &=\frac{d\langle \hat{H}(\vec{\theta})\rangle}{d\theta_k}\Big{|}_{\vec{\theta} =0} =  \langle \Psi_0| \hat{U}^{\dagger}(0)\hat{H}\frac{d \hat{U}(0)}{d\theta_k}|\Psi_0\rangle+\langle \Psi_0|\frac{d \hat{U}^{\dagger}(0)}{d\theta_k}\hat{H} \hat{U}(0)|\Psi_0\rangle\nonumber\\
&= \langle \Psi_0| \hat{H} \hat{\sigma}_k|\Psi_0\rangle+\langle \Psi_0| \hat{\sigma}_k^\dagger \hat{H} \hat|\Psi_0\rangle=2\text{Re}\langle \Psi_0|\hat{H}\hat{\sigma}_k|\Psi_0\rangle,
\end{align}
and
\begin{align}\label{Akm}
A_{km} = \frac{d^2\langle \hat{H}(\vec{\theta})\rangle}{d\theta_k d\theta_m}\Big{|}_{\vec{\theta} =0} &= \langle \Psi_0| \hat{U}^{\dagger}(0)\hat{H} \frac{d^2 \hat{U}(0)}{d\theta_k d\theta_m} |\Psi_0\rangle +2 \langle \Psi_0| \frac{\hat{U}^{\dagger}(0)}{ d\theta _k}\hat{H}\frac{d \hat{U}(0)}{d\theta_ m}|\Psi_0\rangle\nonumber\\
&+    \langle \Psi_0| \frac{d^2 \hat{U}^{\dagger}(0)}{d\theta_k d\theta_m}\hat{H}  \hat{U}(0)|\Psi_0\rangle \nonumber\\
&=  2 \langle \Psi_0|\hat{\sigma}^{\dagger}_k \hat{H} \hat{\sigma}_m|\Psi_0\rangle + 2 \text{Re}\langle \Psi_0| \hat{H} (\hat{\sigma}_k \hat{\sigma}_m)|\Psi_0\rangle.
\end{align}
Note that the real part is not necessary in most calculations, since the wavefunction is usually expanded in terms of real coefficients; this can change in the presence of a magnetic field or with spin-orbit coupling.
The ordering of the operators, however,  is important. Note that in
Eq.~(\ref{Akm}), we have used the following notation:
\begin{align}
    (\hat{\sigma}_k \hat{\sigma}_m)= 
\begin{cases}
    \hat{\sigma}_k \hat{\sigma}_m,& \text{if } m\geq k\\
      \hat{\sigma}_m \hat{\sigma}_k      & \text{otherwise}
\end{cases}.
\end{align}

The ordering of the UCC factors matters in the second derivative matrix because of second term in Eq.~(\ref{Akm}). The de-excitation operator may also apply when the first operator can be de-excited. Since $\hat{\sigma}_k|\textrm{HF}\rangle$ and $\hat{\sigma}_k\hat{\sigma}_m|\textrm{HF}\rangle$ are both single determinants, the above expressions for $A$ and $b$ are just many-body Hamiltonian matrix elements in the Hartree-Fock basis and with respect to many-body product states (determinants). This means one can perform an initial minimization, about the point where all angles are zero, from the Hamiltonian, expressed in the Hartree-Fock basis and extended to include matrix elements for all states that are required in the $b_k$ vector and the $A_{km}$ matrix.

To minimize the energy with respect to the angles, we  take the derivative of Eq.~(\ref{tot_energy}) about the point where each $\theta_i=0$ and set the derivative to zero in order to find the minimum energy.
We have
\begin{align}\label{dev}
\frac{d\langle \hat{H} (\vec{\theta})\rangle}{d\theta_i}\Big |_{\vec{\theta}=0} = b_i + \sum_j A_{ij} \theta_j = 0
\end{align}
In matrix form, the angles that minimize the energy are the solution of the following system of linear equations:
\begin{align}\label{linear_system}
 \mathbf{A} \cdot \vec{\theta} = -\vec{b}.
\end{align}
Angles that minimize the energy can be obtained either by inversion of  the matrix $\mathbf{A}$, or by solving linear system Eq.~(\ref{linear_system}) by row-reduction (or by using sparse-matrix iterative techniques).

The correlation energy, which is the difference between the total energy and the Hartree-Fock energy becomes
\begin{align}
E_{corr} = \langle \hat{H}(\vec{\theta}) \rangle - \langle \hat{H}(0) \rangle = \vec{b}\cdot\vec{\theta}_{min} + \frac{1}{2}\vec{\theta}^{T}_{min} \cdot \mathbf{A} \cdot \vec{\theta}_{min}.
\end{align}
Note how this calculation is quite straightforward and simple to carry out, it just requires generating Hartree-Fock matrix elements and solving the linear matrix equation.

\subsection{Exact UCC Reference}\label{tree_reference}
For this quadratic expansion to be accurate, it is necessary for the angles to be small. This is not the case for strongly correlated systems. To extend this scheme to correlated systems, we can treat factors with large angles more carefully. After separating factors with large angles $\{\hat{\sigma_l}\}$ and small angles $\{\hat{\sigma_s}\}$ into two groups, UCC factors with large angles are used to construct an exact UCC reference wave-function and angles are optimized:
\begin{align}\label{ucc_ref}
    |\Psi^{\textrm{UCC}}_0\rangle = \prod_l e^{\theta_l\hat{\sigma_l}} |\textrm{HF}\rangle. 
\end{align}
This reference state naturally contains more than one determinant, and its contribution to the correlation energy can be calculated as
\begin{align}
    E_{0}^{corr} = \langle \Psi^{\textrm{UCC}}_0 | \hat{H}|\Psi^{\textrm{UCC}}_0\rangle -E_{\textrm{HF}}.
\end{align}

In the second step, angles for all UCC factors are expanded to second order, as described in last section.  The difference is that, for factors used in the UCC reference, their angles are expanded with respect to their optimized values, instead of around zero. The presence of these nonzero angles in the reference state affects the computation of the $b$ vector and the $A$ matrix for all of their elements. The total energy after the quadratic expansion becomes, in this case,
\begin{align}
\langle \hat{H} (\vec{\theta})\rangle = \langle \Psi^{\textrm{UCC}}_0| \hat{H} | \Psi^{\textrm{UCC}}_0 \rangle + \sum_{k} b_k \theta_k + \frac{1}{2}\sum_{k,m} A_{km}\theta_k\theta_m + \mathcal{O}(\theta^3)
\end{align}

Again, $b_k$ and $A_{km}$ can be obtained by taking the corresponding derivatives. But, the two groups of UCC factors, corresponding to large and small angles, have different expressions for the derivatives. For the vector $b_k$, if $k$ is within the group of small angles, elements can be calculated simply by replacing the reference state $|\Psi_0\rangle$ in Eq.~(\ref{b_vec}) by $|\Psi_0^{\textrm{UCC}}\rangle$ as defined in Eq.~(\ref{ucc_ref}); otherwise, $b_k$ can be calculated as a derivative of UCC wave-function, by inserting Eq.~(\ref{operator_identity_2}) into Eq.~(\ref{ucc_first_derivative}). We have an exact expression for the derivative of each UCC factor. Furthermore, we know exactly where to place the derivative operator within the calculation. The matrix $A_{km}$ can be obtained in a similar way---if two factors $k,m$ are both small angle factors, then Eq.~(\ref{Akm}) can be employed with $|\Psi_0^{\textrm{UCC}}\rangle$ replacing  $|\Psi_0\rangle$ . If one or two of $k,m$ are large angle factors, derivatives with respect to $|\Psi_0^{\textrm{UCC}}\rangle$ need to be taken, just like the case for $b_k$, already discussed. It is a simple exercise to work out the exact formulas (we do not write them out here).

The energy minimization is again carried out to find $\theta_{min}$ for both large and small angles. Then the large angles used to construct the UCC reference wave-function are updated by $\theta'_l = \theta_l + \theta_l^{min}$. These $\theta'_l$ values are then used to construct the new UCC reference wavefunction, which is used for the new quadratic angle expansion. This procedure is iterated until the calculation has converged to a fixed point for both the large and the small angles. The total correlation energy now has three contributions:
\begin{align}
E_{corr} =   E_{0}^{corr}  + \vec{b}\cdot\vec{\theta}_{min} + \frac{1}{2}\vec{\theta}^{T}_{min} \cdot \mathbf{A} \cdot \vec{\theta}_{min}.
\end{align}

The computational procedure described above can be carried out as a hybrid quantum-classical algorithm. The preparation of the UCC reference state needs to be performed on a quantum computer. Then the quantities $b_k$ and $A_{km}$ are measured with respect to the prepared UCC reference state. As we will show later, only a small number of factors are required in the UCC reference state, which greatly reduces the circuit depth on a quantum computer. As a trade off, the $b_k$ and $A_{km}$ require many more measurements. But of course, even a standard VQE procedure will require calculating at least something like the $b_k$ for a gradient-based minimizer, or to choose the best operator to pick from an operator pool. If this tradeoff is worthwhile requires a more detailed analysis. But, note that there are likely some measurements that can be replaced by cubic expansion about $\vec{\theta}=0$, because they might not depend strongly on the large angles. This analysis can be performed on a classical computer prior to the quantum computation and can greatly reduce the required measurements.

One potential issue in this procedure is the appearance of an instability in the minimization step for  strongly correlated cases. The origin of this instability is that the inverse of $\mathbf{A}$ may be ill-defined (due to zero or near zero eigenvalues). To overcome this, we use a pseudo-inverse of $\mathbf{A}$, which is constructed as follows. First, $\mathbf{A}$ is diagonalized via a similarity transformation with the matrix $Q$: $\mathbf{A} = \mathbf{Q} \mathbf{\Lambda} \mathbf{Q}^{-1}$. Then, a cutoff $\epsilon$ is applied to the eigenvalues included in $\Lambda$. If an eigenvalue is smaller than $\epsilon$, the corresponding diagonal element of $\Lambda^\prime$ is set to zero. The reduced set of eigenvalues is placed in the diagonal matrix $\Lambda'$, and the pesudoinverse of  $\mathbf{A}$ is calculated as $\mathbf{A}^{-1}\approx \mathbf{Q} \mathbf{\Lambda'}^{-1} \mathbf{Q}^{-1} $, where the terms in $\bf{\Lambda}^\prime$ that were set equal to zero are \textit{not} inverted in computing the pseudoinverse---they remain set equal to zero. Thisapproach is the same as a singular-value decomposition of a matrix, relative to a cutoff $\epsilon$. Then, the angles that minimize the energy are found from $\theta_{min} = -\mathbf{A}^{-1}
\cdot \vec{b} $. 



\section{Results and Discussion}

\subsection{Near equilibrium}

\begin{figure}
\centering
\includegraphics[width=0.85\linewidth,clip]{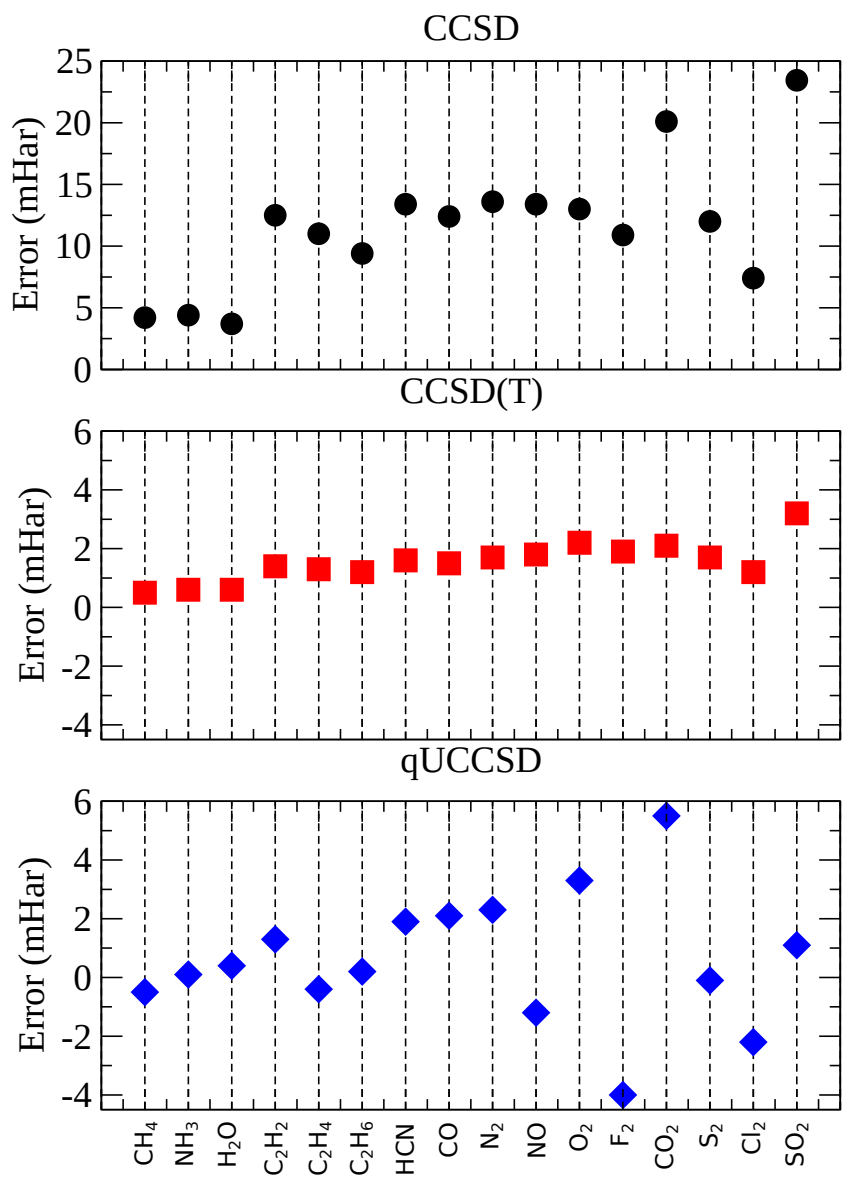}
\caption{ Error, (E-E$_{FCI}$), from CCSD, CCSD(T) and qUCCSD calculations on a subset of Gaussian-2 molecules. The basis set is {\it ccpvdz}. O$_2$ and S$_2$ are spin triplets, and NO is a spin doublet. All others are spin singlets. } \label{tt}
\end{figure}

We implemented this quadratic UCC (qUCC) method using integrals generated by {\it PySCF}. We performed calculations on a set of small molecules with the  {\it ccpvdz} basis set. To benchmark the calculations, we use almost exact energies from the semistochastic heat-bath configuration interaction calculation as a reference.\cite{yao_2020} The results are summarized in Fig.~\ref{tt}.

While an exact UCC calculation is always variational, the qUCCSD approximation need not have its energy bounded by the full configuration interaction (FCI) from below, because of the potential error from the truncated Taylor expansion. This means it need not be strictly variational. For several molecules in Fig.~\ref{tt}, the errors are indeed negative. However, if we compare the absolute error, qUCCSD is more accurate than CCSD, which has similar computational scaling. Comparing to CCSD(T), qUCCSD is not as accurate, but is not expected to be either. What is surprising is that qUCCSD is consistently not that far away from CCSD(T).
 
\subsection{Chemical Bond Stretching}
Thje quadratic expansion of the unitary coupled cluster theory using HF ground state as reference, as described earlier, shows many similar characteristics with the older XCC approaches \cite{Bartlett1989}, and other linearized coupled-cluster methods \cite{Bartlett2007} developed by Rodney Bartlett and his colleagues. But even at second order, the derivatives from a factorized form of the UCC depend on the ordering, while for a conventional UCC, they do not. The common weakness of the XCC methods is that they diverge when near-degeneracy is present in the HF states, so they cannot treat level crossings or near level crossings in the potential energy surface. This often occurs for bond stretching and breaking situations. One way to overcome this difficulty for the quadratic expansion of the factorized form of the unitary coupled cluster theory is to use multi-determinant reference states, as detailed in Section.~\ref{tree_reference}.

\begin{figure}
\centering
\includegraphics[width=0.85\linewidth,clip]{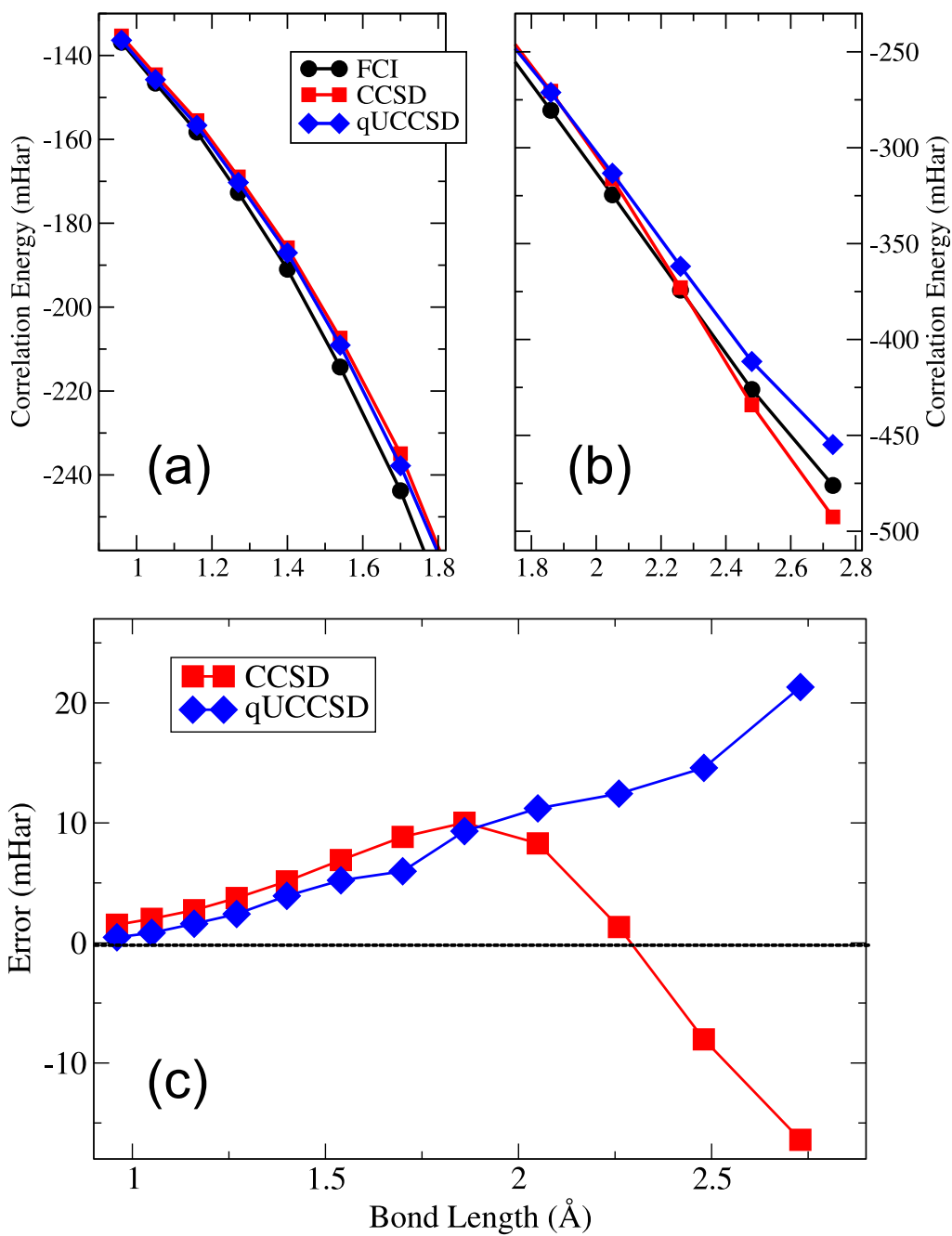}
\caption{ Panel (a) and panel (b): total correlation energies  for the water molecule as function of the H-O bond length. Here, we plot results for FCI (black circles), CCSD (red squares) and qUCCSD (blue diamonds) with a UCC reference state (that contains 28 exact UCC factors). Note that the ranges of the plots differ in the two panels.  Panel (c) error of the CCSD (red squares) and qUCCSD (blue diamonds) results relative to the FCI as a function of the H-O bond length. Note how the CCSD calculation becomes nonvariational at strong coupling.  } \label{ucc_reference}
\end{figure}

To demonstrate this method, we performed calculations for the bond stretching of a H$_2$O molecule with the 6-31G basis set. In total, this system has 74 single and 2240 double excitations. We first did a geometry optimization, and then optimized the H-O bond length. It becomes 0.96\AA ~and the optimized H-O-H angle is 104.5$^{\circ}$. To study the situations of bond stretching, we gradually increase the bond from its equilibrium length to 2.74\AA, while keeping the H-O-H angle fixed. Standard quantum chemistry methods inclduing full configuration interaction (FCI) and coupled-cluster singles and doubles (CCSD) were carried to test against the qUCCSD with exact UCC reference states. Here all the qUCCSD calculations were performed with 28 factors of large angles and $\epsilon = 0.1$ for pseudo-inverse procedure. Those 28 large angles factors are determined from initial MP2 amplitudes. Results can be found in Fig.~\ref{ucc_reference}.

As expected, CCSD correlation energies are close to the exact FCI results when the molecular geometries are close to equilibrium. But, CCSD has difficulties when away from equilibrium; CCSD violates the variational principle and gives lower energies then FCI when O-H bond length is larger then 2.2 \AA. We also find that qUCCSD with UCC reference usually gives better results than CCSD. Close to equilibrium, qUCCSD gives similar results as CCSD, and when far equilibrium, qUCCSD with UCC reference states has better behaviour then CCSD, since it never violates variational principle.

\section{Conclusion}

One of the main issues slowing down the ability for quantum computers to show an advantage over classical computers is that current hardware can only run codes that are quite short in circuit depth. In this situation, the only advantage will occur due to the larger memory that a quantum computer has in storing quantum states. The algorithm discussed here, denoted qUCC, is a methodology that will allow quantum computers to aid in determining the electronic structure of complex molecules much sooner, because it trades off circuit depth for additional measurements. If those additional measurements can be carried out, then we might see a quantum advantage sooner than later.

To illustrate how this methodology works, we showed that the number of exact UCC factors needed in an electronic structure calculation on a quantum computer can be greatly reduced from that of a standard approach. Our test case (bond stretching of water molecule) shows better results can be achieved with only a small fraction of factors (28 versus 2314) when constructing the wavefunction. Our result compares the qUCC approach to a CC approach, but results would be similar for a comparison to a standard UCC approach as well. Our algorithm is one way to utilize low-depth quantum circuits to treat molecules with large basis sets, which has been a major obstacle for applying quantum computing to quantum chemistry. 

\begin{acknowledgement}
 JC and HPC are supported by the Department of Energy, Basic Energy Sciences as part of the Center for Molecular Magnetic Quantum Materials, an Energy Frontier Research Center under Award No. DE-SC0019330. JKF is supported from the National Science Foundation under grant number CHE-1836497. JKF is also funded by the McDevitt bequest at Georgetown University. This research used resources of the National Energy Research Scientific Computing Center (NERSC), a U.S. Department of Energy Office of Science User Facility operated under Contract No. DE-AC02-05CH11231, and University of Florida Research Computing systems.
 We also acknowledge useful discussions with Rodney Bartlett, Garnet Chan, Joseph Lee, John Staunton, Cyrus Umrigar, Luogen Xu and Dominika Zgid.
\end{acknowledgement}

\bibliography{Collection}

\end{document}